\documentclass[
    ,final            % use final for the camera ready runs
  ]
  {aipproc}

\layoutstyle{6x9}

\begin{document}

\title{Simbol-X Background Minimization: Mirror Spacecraft Passive Shielding Trade-off Study}


\author{V. Fioretti}{
  address={INAF/IASF -- Bologna, Via Gobetti 101, 40129 Bologna, Italy},
  altaddress={Dipartimento di Astronomia, Universit\`a di Bologna, via Ranzani 1, 40127 Bologna, Italy}
}

\author{G. Malaguti}{
  address={INAF/IASF -- Bologna, Via Gobetti 101, 40129 Bologna, Italy},
}

\author{A. Bulgarelli}{
  address={INAF/IASF -- Bologna, Via Gobetti 101, 40129 Bologna, Italy},
}

\author{G.G.C. Palumbo}{
  address={Dipartimento di Astronomia, Universit\`a di Bologna, via Ranzani 1, 40127 Bologna, Italy}
}

\author{A. Ferri}{
  address={Thales Alenia Space, Corso Marche 41, 10146 Torino, Italy}
}

\author{P. Attin\`a}{
  address={Thales Alenia Space, Corso Marche 41, 10146 Torino, Italy}
}

\begin{abstract}
The present work shows a quantitative trade-off analysis of the Simbol-X Mirror Spacecraft (MSC) passive shielding, in the phase space of the various parameters: mass budget, dimension, geometry and composition. A simplified physical (and geometrical) model of the sky screen, implemented by means of a GEANT4 simulation, has been developed to perform a performance-driven mass optimization and evaluate the residual background level on Simbol--X focal plane.
\end{abstract}

\maketitle

%%%%%%%%%%%%%%%%%%%%%%%%%%%%%%%%%%%%%%%%%%%%
%% MAINMATTER
%%%%%%%%%%%%%%%%%%%%%%%%%%%%%%%%%%%%%%%%%%%%

\section{Introduction}
The scientific requirements and the advanced design of Simbol--X \cite{ferrando2005} imply the necessity of great care in the minimization of the background radiation at high energy. This is canonically achieved with a detailed characterization of the expected background events \cite{tenzproc} , which will also be a key input for the Simbol-X instrument scientific calibration \cite{malaproc}.
In this context, the background events can be broadly divided into two main categories: (a) the ones due to diffuse CXB photons, and (b) the hadronic component originated by prompt and delayed events caused by high energy particles.
The minimization of the photonic component is generally achieved by shielding the detector aperture to the unfocused photons through an high Z material.
\\
In this work a quantitative trade-off analysis of the passive shielding on board the Simbol--X Mirror Spacecraft (MSC) is addressed considering both its geometry and composition, with the aim of evaluating the impact of the required mass budget and design on the shielding efficiency and photonic background level.

\section{Simbol-X MSC passive shielding design}
\begin{figure}[!h]
  \includegraphics[height=.225\textheight]{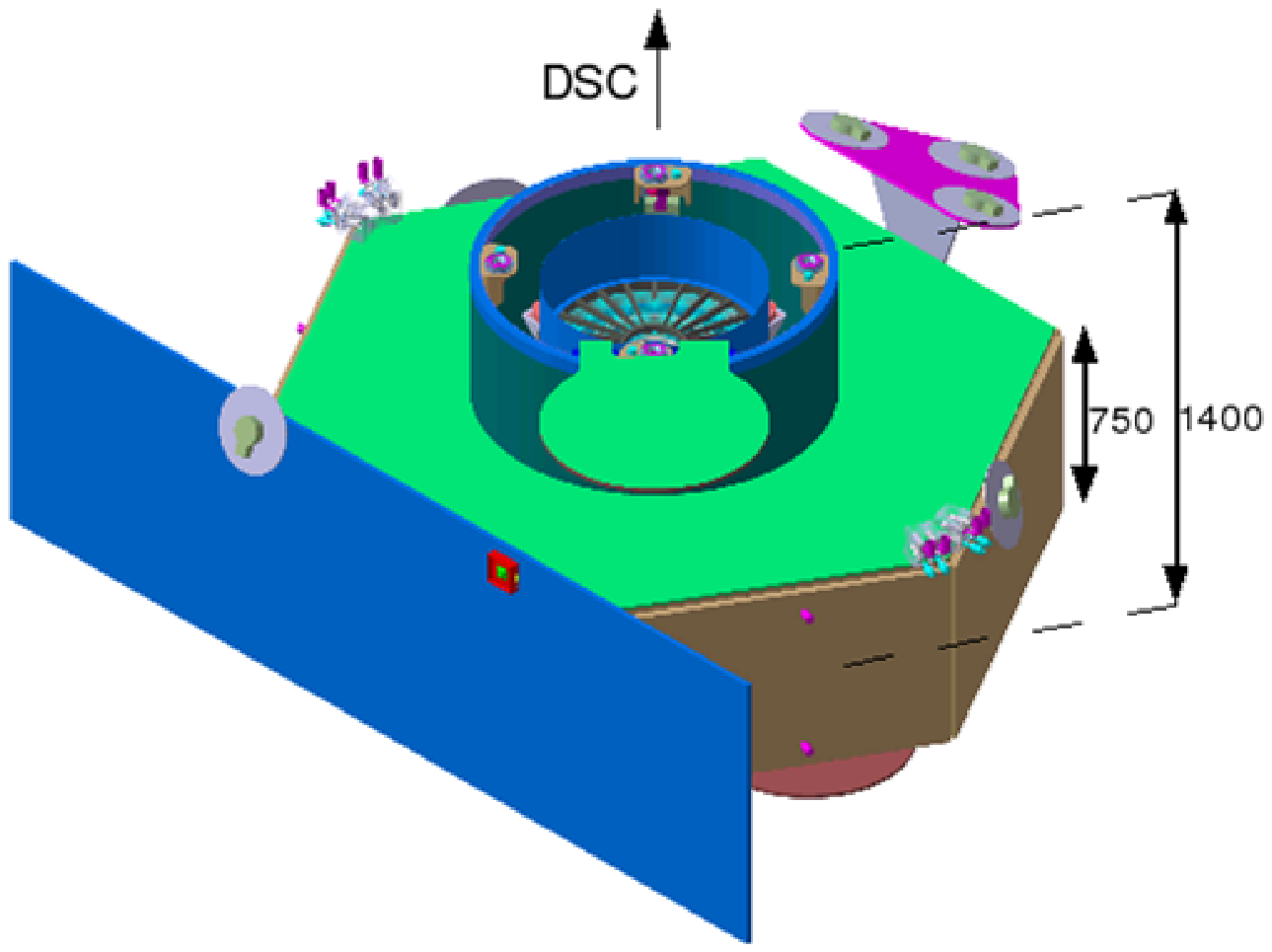}
  \includegraphics[height=.225\textheight]{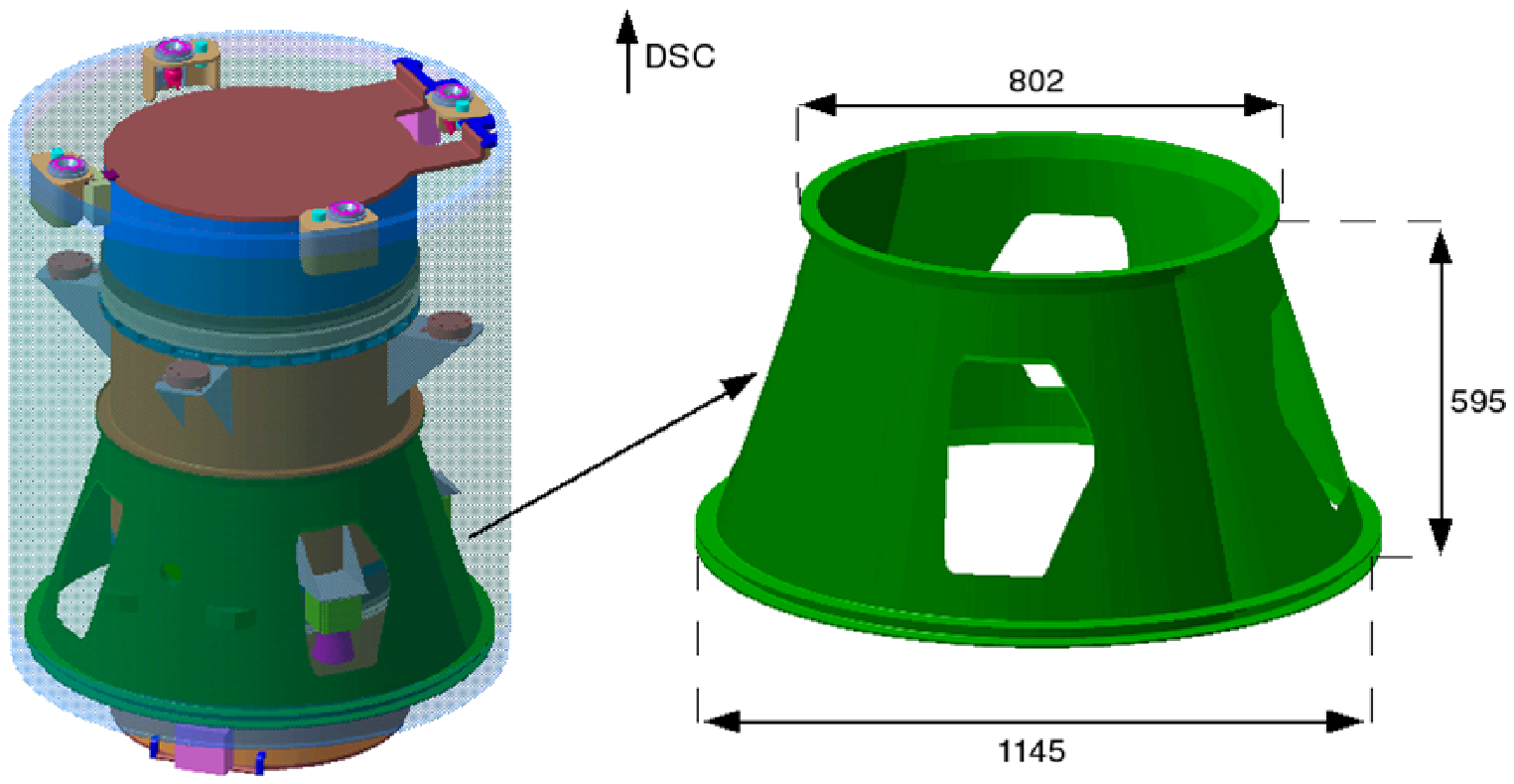}
  \caption{3D view of the MSC geometry (left panel), with zoom on the thrust cylinder (right panel).}
\label{fig:thales}
\end{figure}

Since the formation flight architecture avoids the possibility of using the canonical telescope "tube" connecting the mirror to the focal plane unit, Simbol--X passive shielding system consists of two main parts \cite{mala2005}: the collimator tube placed on top of the focal plane and an hexagonal passive shield (sky screen) around the mirror module (MM).
\begin{figure}[!h]
  \includegraphics[height=.2\textheight]{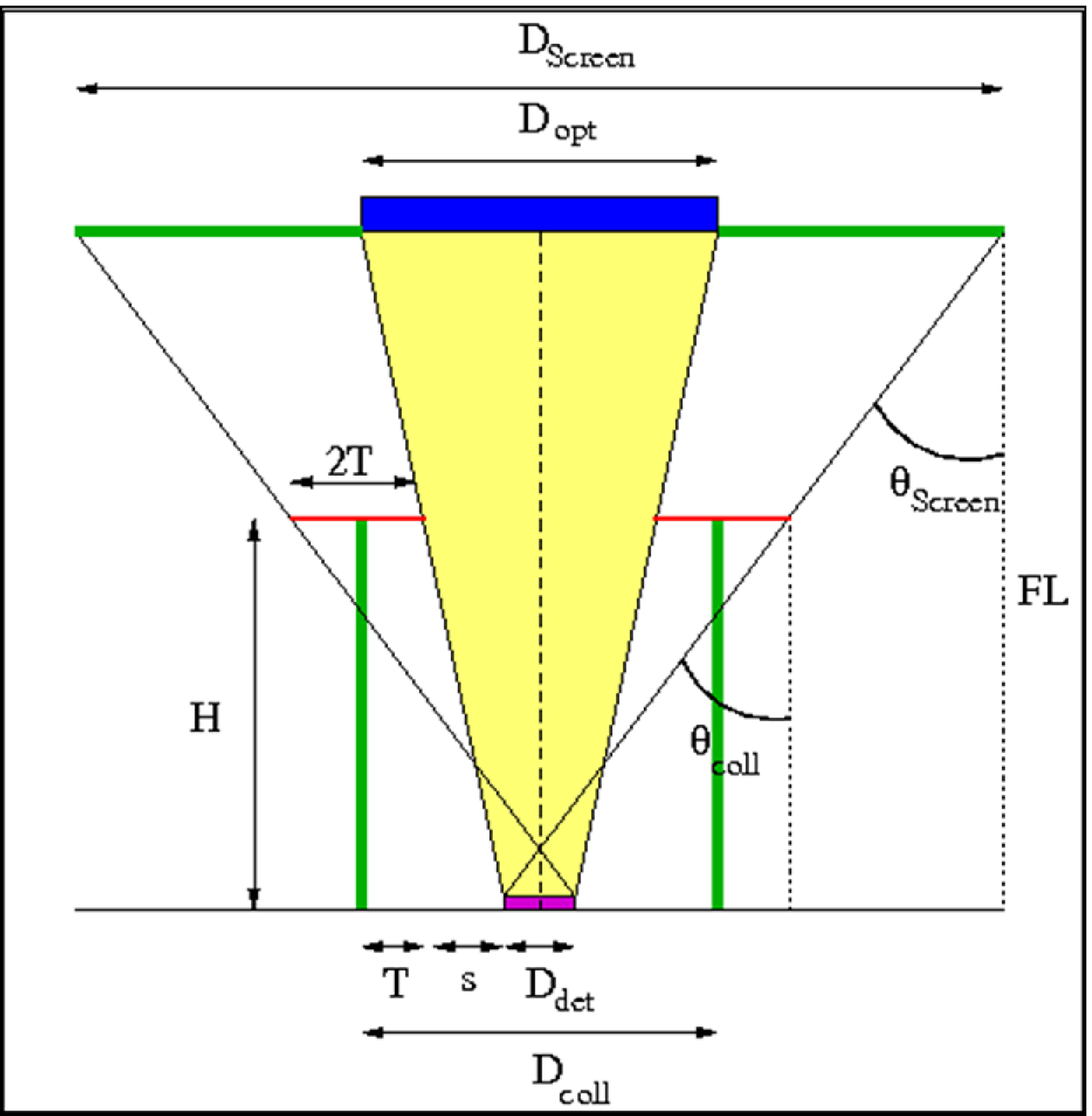}
  \includegraphics[height=.2\textheight]{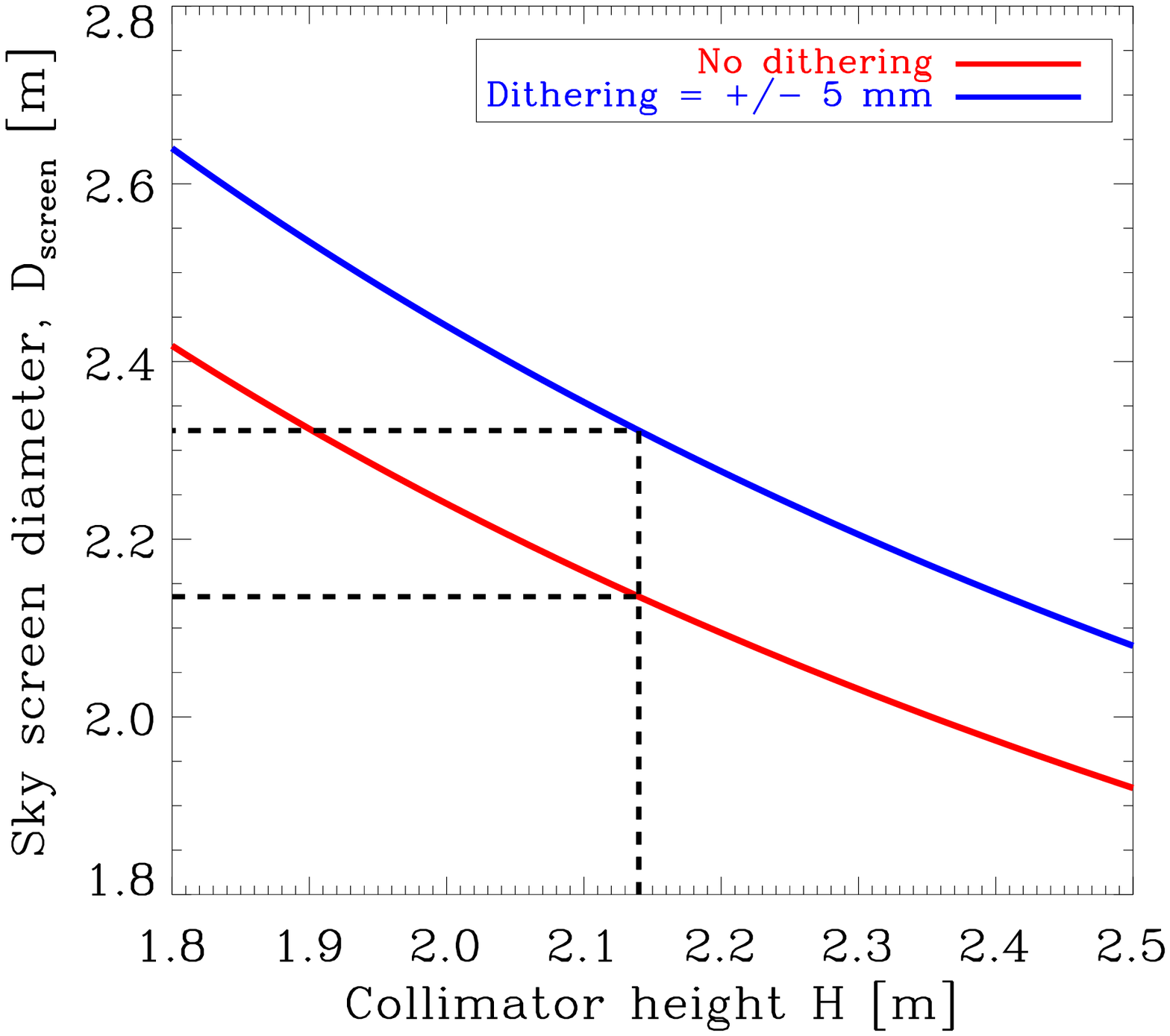}
  \caption{(\textit{Left}): Simplified Simbol--X passive shielding geometry for a $\pm$ T lateral dithering. (\textit{Right}): Sky screen diameter in function of the collimator height for a perfect shielding.}
\label{fig:per_shield}
\end{figure}
The passive shielding design must account for the formation flight relative motion, since a lateral dithering between the spacecrafts results in both a collimator walls vignetting effect and a detector opening angle to the unfocused CXB photons.
If we assume a collimator height of 2.14 m and a $\pm$5 mm dithering, the sky screen diameter required for a perfect shielding is 2.32 m (see Fig. \ref{fig:per_shield}).
\\
The MSC design\footnote{The MSC baseline configuration and figures described in this section have the only purpose of evaluating the MSC passive shielding geometry and are courtesy of AF (Simbol-X payload meeting, 04/2008)} is characterized by an hexagonal platform (Fig. \ref{fig:thales}, left panel), with the MM placed inside the thrust cylinder (Fig. \ref{fig:thales}, right panel). The adapter, the cone trunk structure, is the interface between the MM and the thrust cylinder.
The presence of the thrust cylinder around the optics hampers a complete shielding around the MM. This region is shielded by coating, using the same passive material of the sky screen, the adapter except for the holes, the contribution of which to the background level must be evaluated.
The MSC passive shielding geometry will thus foresee the sky screen (an hexagonal plate covering the region around the thrust cylinder) and the adapter cover (on the DSC side). Since the sky screen flat-to-flat diameter is 2.38 m, the Simbol--X focal plane is shielded for a dithering up to $\pm$5 mm.
\\
The selected materials for the MSC passive shielding main absorber are Pb, W and Ta. Since the main absorber fluorescence lines fall within Simbol--X energy range, a grading (Sn + Cu + Al + C) is added.

\section{MSC passive shielding background evaluation}
\begin{figure}[!h]
  \includegraphics[height=.23\textheight]{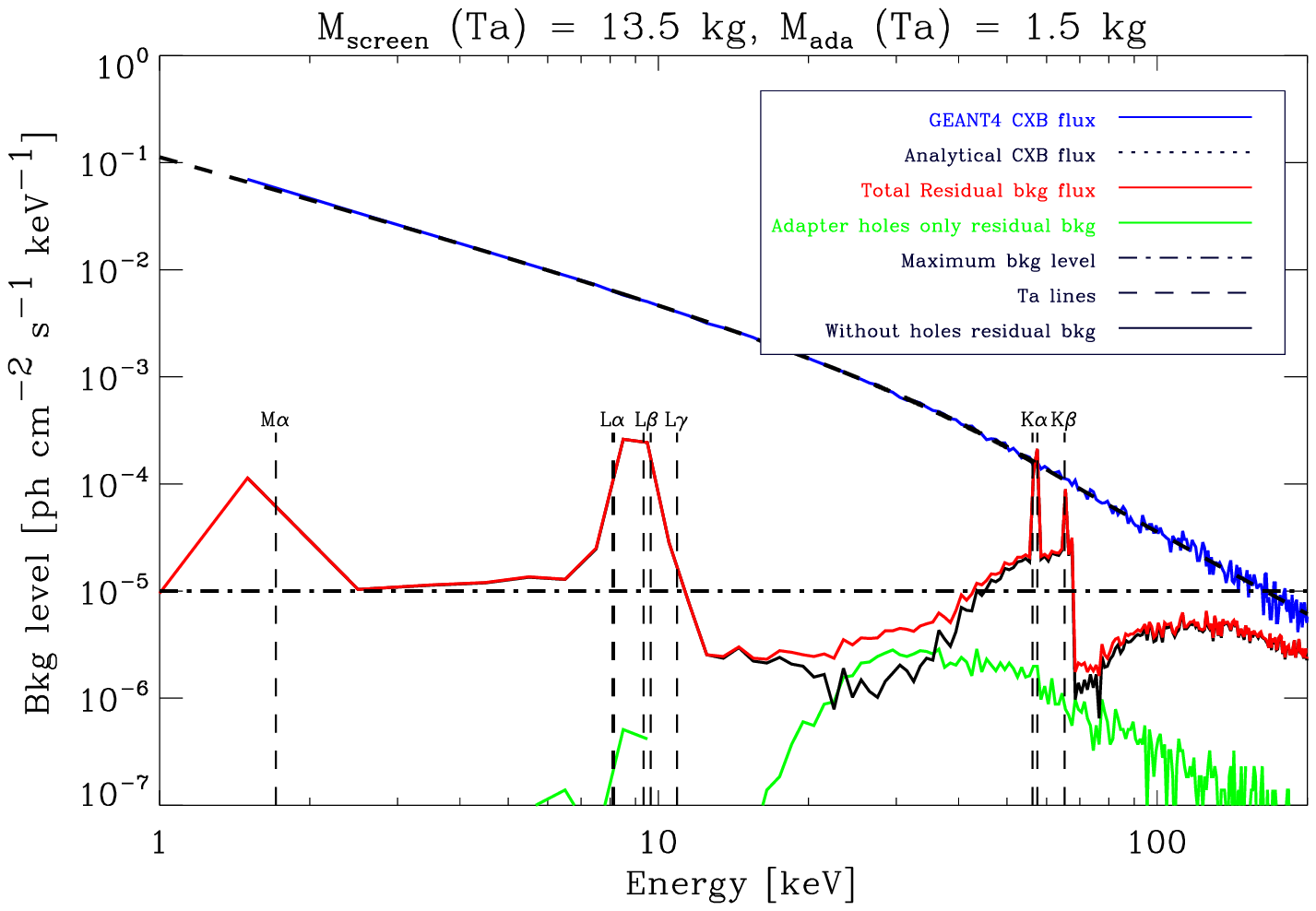}
  \includegraphics[height=.23\textheight]{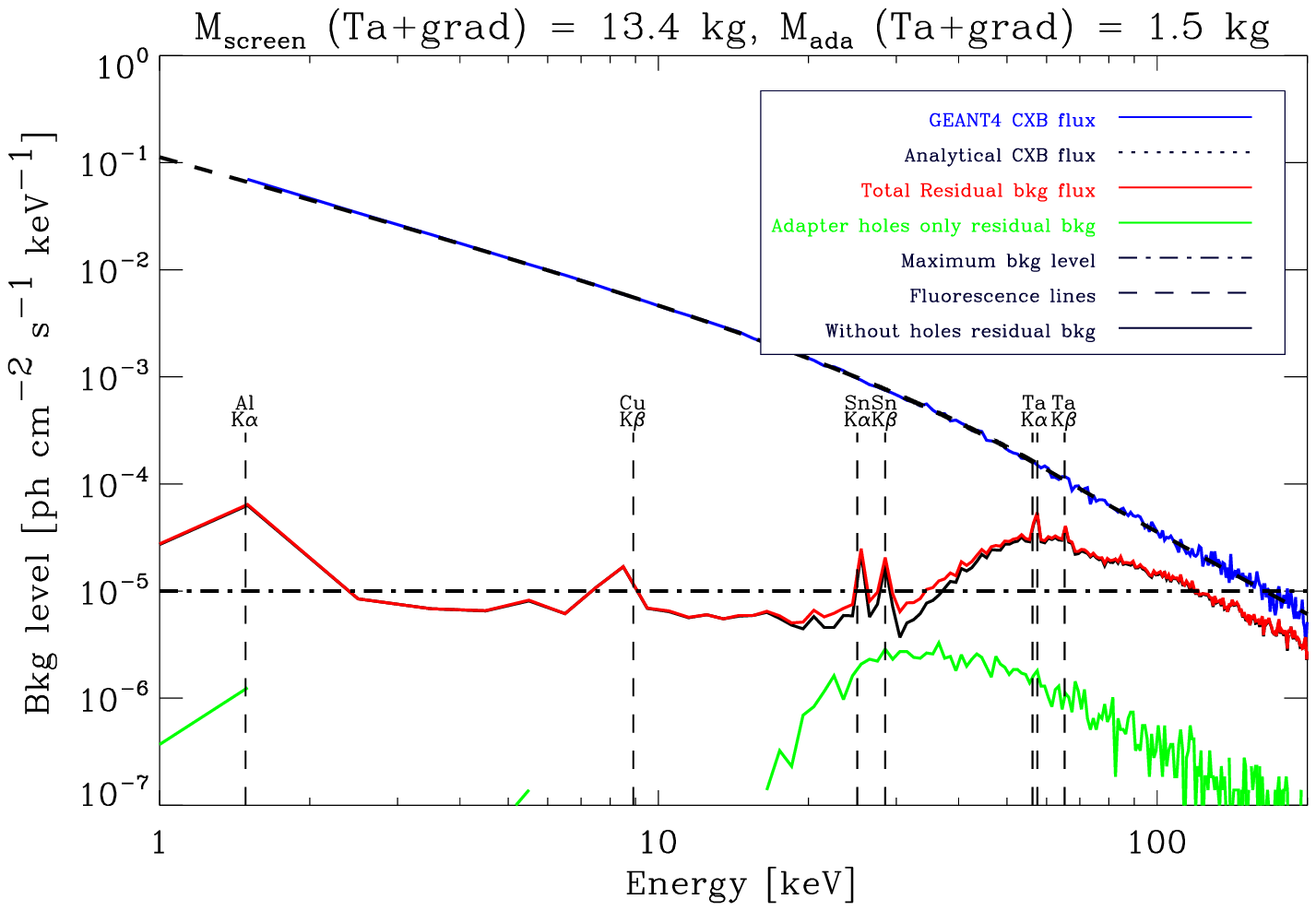}
  \caption{MSC passive shielding residual background spectra for an only main absorber composition (left panel) and a Ta plus grading composition (right panel).}
\label{fig:res}
\end{figure}

In order to evaluate the photonic residual background level (due to leaking or fluorescence photons) on Simbol--X focal plane, a GEANT4 \cite{Geant2003} based Monte Carlo simulation of the MSC passive shielding interaction with the CXB flux \cite{Gruber1999} has been implemented. 
The absorption effect of the MSC is simulated with a 3.5 mm thick Al layer on the backside of the shielding while the thermal blankets are represented by a 1 mm thick C layer.
\\
The spectrum of the MSC passive shielding residual background (Fig. \ref{fig:res}) is evaluated for a 15 kg total mass budget and a Ta main absorber (with and without the grading layers).
\\
The horizontal dashed-dotted line shows the maximum accepted residual background level ($10^{-5}$ cts cm$^{-2}$ s$^{-1}$ keV$^{-1}$), given by the Simbol--X scientific requirements.
\\
The resulting background spectra indicate that the fluorescence emission for an only main absorber composition is well above the limit and the grading layers are required.\section{Conclusions}
The main result of the present work is that a 15 kg total mass budget allows for an average residual background flux within the requirement, with the thrust cylinder leakage component being efficiently absorbed by the payload and thermal covering. 
\\
Although the grading layers absorb about 90\% of the main absorber fluorescence emission, K$\alpha$ and K$\beta$ lines are still present in the residual background spectrum. These might be minimized by a further grading layers optimization.

%%%%%%%%%%%%%%%%%%%%%%%%%%%%%%%%%%%%%%%%%%%%%%%%
%% BACKMATTER
%%%%%%%%%%%%%%%%%%%%%%%%%%%%%%%%%%%%%%%%%%%%%%%%

%%%%%%%%%%%%%%%%%%%%%%%%%%%%%%%%%%%%%%%%%%%%%%%%
%% The bibliography can be prepared using the BibTeX program or
%% manually.
%%
%% The code below assumes that BibTeX is used.  If the bibliography is
%% produced without BibTeX comment out the following lines and see the
%% aipguide.pdf for further information.
%%
%% For your convenience a manually coded example is appended
%% after the \end{document}
%%%%%%%%%%%%%%%%%%%%%%%%%%%%%%%%%%%%%%%%%%%%%%%%

%%%%%%%%%%%%%%%%%%%%%%%%%%%%%%%%%%%%%%%%%%%%%%%%
%% You may have to change the BibTeX style below, depending on your
%% setup or preferences.
%%
%%
%% For The AIP proceedings layouts use either
%%%%%%%%%%%%%%%%%%%%%%%%%%%%%%%%%%%%%%%%%%%%

%\bibliographystyle{aipproc}   % if natbib is available
%\bibliographystyle{aipprocl} % if natbib is missing

%%%%%%%%%%%%%%%%%%%%%%%%%%%%%%%%%%%%%%%%%%%
%% You probably want to use your own bibtex database here
%%%%%%%%%%%%%%%%%%%%%%%%%%%%%%%%%%%%%%%%%%%
%\bibliography{sample}

\end{document}